\documentclass[11pt]{article}
\usepackage{color}
\usepackage{subfigure}
\usepackage{amsmath}
\usepackage{amssymb}
\usepackage{graphicx,color}
\usepackage{cite}
\usepackage{enumerate}
\usepackage{amsthm}
\usepackage{amsfonts,mathrsfs}
\usepackage{geometry}
\usepackage{ifthen}
\usepackage{graphicx}
\usepackage{epstopdf}
\usepackage{float}
\usepackage{bm}
\usepackage[caption=false]{subfig}
\usepackage[utf8]{inputenc}
\usepackage{graphicx}
\usepackage{epsf}
\usepackage{CJK}

\begin{document}

\title{\bf Coherence and decoherence in generalized Shor's algorithm}

\vskip0.1in

\author{
\begin{minipage}{1.0\textwidth}
\centering {\small Linlin Ye$^1$
,Zhaoqi Wu
$^1$\thanks{Corresponding author. E-mail:
wuzhaoqi\_conquer@163.com}, Nanrun Zhou
$^{2,3}$}\\
{\small\it 1. Department of Mathematics, Nanchang University,
Nanchang 330031, China}\\
{\small\it 2. School of Electronic and Electrical Engineering, Shanghai University of Engineering Science, Shanghai 201620, China}\\
{\small\it 3. Department of Electronic Information Engineering, Nanchang University, Nanchang 330031, China}
\end{minipage}
}

\date{}
\maketitle

\noindent {\bf Abstract} {\small }\\
Quantum coherence constitutes a fundamental physical mechanism
essential to the study of quantum algorithms. We study the coherence
and decoherence in generalized Shor's algorithm where the register
$A$ is initialized in an arbitrary pure state, or the combined register
$AB$ is initialized in a pseudo-pure state, which encompasses the
standard Shor's algorithm as a special case. We derive both the
lower and upper bounds on the performance of the generalized Shor's
algorithm, and establish the relation between the probability of
calculating the order $r$ when the register $AB$ is initialized in a
pseudo-pure state and the one when the register $A$ is initialized in an
arbitrary pure state. Moreover, we study the coherence and
decoherence in noisy Shor's algorithm and give the lower bound of
the probability that we can calculate the order $r$.

\noindent {\bf Keywords}: Shor's algorithm; Coherence; Decoherence; Success probability\\
{\bf PACS}: 03.67.Ac, 03.65.Yz, 03.67.-a

\vskip0.2in

\noindent {\bf 1 Introduction}\\\hspace*{\fill}\\
Quantum coherence represents a fundamental resource in quantum
information processing, derived from the quantum superposition
principle, a concept originally formulated to explain interference
phenomena in light\cite{SEC,GRJ,SMO}. The rigorous quantification of
coherence has been systematically formalized through comprehensive
theoretical frameworks \cite{TB,YXD}, which have subsequently
facilitated the development of diverse coherence measures with
distinct operational
interpretations\cite{XJCM,BKFS,YCS,ZXN,SLH,WZZ}. As
a quantifiable resource closely connected to other ones, coherence
has many applications in other fields
\cite{WinterYang2016,LiuFan2023,Huang2024,StreltsovRMP,HuPhysRep}.
Significant progress has been made in systematic investigation of
coherence manipulation, evolving from fundamental studies on
coherence transformations under incoherent operations to
sophisticated protocols for multi-mode quantum systems
\cite{Dushuanping1,Dushuanping2,Dushuanping3,Dushuanping4}.
Furthermore, quantum coherence has been extended from orthonormal
bases-based to POVM-based\cite{Bischof2019,Guo2023,Melo2025}.

Fisher information stands as a fundamental concept
in quantum metrology \cite{HAS,BSL,GVS}. Utilizing operator monotone
metrics, Hansen\cite{HFM} introduced a class of quantum Fisher
information known as metric-adjusted skew information, which
encompasses notable examples such as the Wigner-Yanase skew
information and the quantum Fisher information based on the
symmetric logarithmic derivative. Owing to its versatility and
theoretical importance, metric-adjusted skew information has found
widespread applications across various domains, including
uncertainty relations \cite{LSH,RRP} and asymmetry \cite{TRS}. Sun
\cite{SYL} further demonstrated its utility by quantifying the
coherence of quantum states relative to channels using
metric-adjusted skew information.


Shor's algorithm \cite{PWSP} enables the efficient
factorization of large integers, posing a significant threat to RSA
encryption security. Accordingly, great attention has been paid on
optimizing its circuit and architectural design to reduce qubit
requirements \cite{Beauregard,Haner,Parker}. It has been clarified
that how Shor's algorithm can be implemented in realistic quantum
models to break RSA, underscoring the vulnerability of conventional
public-key cryptosystems\cite{Kute2024}. Despite these advances,
resource estimates indicate that breaking standard 2048-bit RSA with
Shor's algorithm would still require fault-tolerant quantum hardware
with on the order of millions of physical qubits
\cite{Gidney,Gidney2025}. In parallel with large-scale assessments,
proof-of-principle demonstrations have also been reported, including
implementations based on a single photon encoded in a 32-dimensional
quantum state \cite{Weng2024} and Qiskit-based circuit realizations
for factoring small integers \cite{Gharbi2024}.

Quantum coherence has emerged as a critical resource underpinning
the efficacy of quantum algorithms
\cite{FSH,RPF,MMY,CCH,YWF1,YWF2,YWF3,NMTV,Feng}. Studies have
demonstrated that the Deutsch-Jozsa algorithm's precision correlates
directly with recoverable coherence\cite{HMC,JMM}. Extensive
research has been conducted to elucidate the coherence dynamics in
Grover's search algorithm under both ideal and noise-affected
conditions, establishing significant correlations between coherence
measures and algorithmic success probability
\cite{SY,PMQ,MPH,Anand,Shi,Chin,Rastegin1,Rastegin2,Liu,Rastegin3}.
Similarly, coherence establishes both upper and lower bounds on
Shor's algorithm's performance\cite{AFTTE}. More recently, Zhou
showed that the generalized Grover algorithm's success probability
depends not only on oracle queries but also coherence fraction
\cite{ZSQ}. In this work, we investigate quantum
coherence within the framework of generalized Shor's algorithm.

The remainder of this paper is organized as follows.
In Section 2, we review Shor's algorithm and define
coherence and decoherence based on the generalized metric-adjusted
skew information of $\rho$ with respect to a channel $\Phi$.  In
Section 3, we investigate the coherence and decoherence dynamics in
generalized Shor's algorithm. We derive both lower and upper bounds
on the performance of generalized Shor's algorithm when the register
is initialized in an arbitrary pure state, and derive the relation
between the probability of calculating $r$ for an initial arbitrary
pure state in register $A$ and for an initial pseudo-pure state in
register $AB$. In Section 4, we study the coherence
and decoherence in noisy Shor's algorithm. Finally, we summarize the
results in Section 5.

\vskip0.2in

\noindent {\bf 2. Shor's algorithm and coherence quantifiers}

In this section, we recall Shor's algorithm and the coherence
measures that will be used in this paper.

Given an odd composite positive integer $N$, the prime factorization problem requires finding its prime factors. It is widely recognized that the factorization of a composite number can be reduced to the problem of order finding. Specifically, for an integer $N$ and a coprime integer $x$ where $x<N$ and $\gcd(x, N) = 1$, the order-finding task involves determining the smallest positive integer $r$ satisfying $x^r \equiv 1 \pmod{N}$. The algorithm employs a quantum system of \( n \) qubits structured into two registers: register $A$ containing $t$ qubits (where $Q = 2^t$) and register $B$. The composite system $AB$ is initialized in the state $|0\rangle_{A}^{\otimes t}|1\rangle_{B}$, establishing the foundation for the quantum computation. The principal steps of Shor's algorithm proceed as follows\cite{PWSP}:\\
(i) Impose Hadamard gate $H^{\otimes t}=\frac{1}{\sqrt{2^t}}\sum_{x,y}(-1)^{xy}|y\rangle\langle x|$ on the qubits in register $A$. This yields the state
\begin{equation}\label{eq1}
|\psi_{1}\rangle=\frac{1}{\sqrt{2^t}}\sum_{j=0}^{2^t-1}|j\rangle_{A}|1\rangle_{B},
\end{equation}
and $|\psi_{1}\rangle$ can be rewritten as $|+\rangle^{\otimes t}_{A}|1\rangle_{B}$.\\
(ii) Consider the unitary operator
$U=\sum_{m=0}^{2^t-1}|m\rangle\langle m|_{A}\otimes U_{B}^{m}$
acting on the state $|\psi_{1}\rangle$, where
$U_{B}|m\rangle_{B}=|xm \bmod  N\rangle_{B}$ and
$U|j\rangle_{A}|m\rangle_{B} = |j\rangle_{A}|x^j m \bmod
N\rangle_{B}$. Then we have
\begin{equation}\label{eq2}
|\psi_{2}\rangle = \frac{1}{\sqrt{2^t}}\sum_{j=0}^{2^t-1}|j\rangle_{A}|x^{j} \bmod N\rangle_{B}.
\end{equation}
The eigenvectors of $U_B$, denoted as $|u_{s}\rangle_B$, are given by
$|u_{s}\rangle_B=\frac{1}{\sqrt{r}}\sum_{a=0}^{r-1} \mathrm{e}^{-\frac{2\pi \mathrm{i}as}{r}}|x^{a} \bmod N\rangle_B$, where $0\leq s\leq r-1$.
These eigenvectors correspond to eigenvalues $\mathrm{e}^{2\pi \mathrm{i} s / r}$, and they satisfy the relation $\frac{1}{\sqrt{r}}$ $\sum_{s=0}^{r-1}|u_{s}\rangle_B=|1\rangle_B$. The state $|\psi_{2}\rangle$ can be rewritten as
\begin{equation}\label{eq3}
U|\psi_{1}\rangle= \frac{1}{\sqrt{2^{t}}} \sum_{j=0}^{2^{t}-1} |j\rangle_{A} U_{B}^{j}\left(\frac{1}{\sqrt{r}} \sum_{s=0}^{r-1} |u_s\rangle_{B}\right) = \frac{1}{\sqrt{2^{t}r}} \sum_{s=0}^{r-1} \sum_{j=0}^{2^{t}-1}\mathrm{e}^{2\pi \mathrm{i}j \frac{s}{r}} |j\rangle_{A} |u_s\rangle_{B}.
\end{equation}
(iii) By applying the inverse Fourier transform $F^\dagger$ to register $A$, we obtain
\begin{align}\label{eq4}
|\psi_{3}\rangle
&=(F^\dagger_A\otimes \mathbf{I}_B)\frac{1}{\sqrt{2^{t}r}} \sum_{s=0}^{r-1} \sum_{j=0}^{2^{t}-1}\mathrm{e}^{2\pi \mathrm{i}j \frac{s}{r}} |j\rangle_{A} |u_s\rangle_{B}\notag\\
&=\frac{1}{\sqrt{r}} \sum_{s=0}^{r-1} \left(F^\dagger_A\frac{1}{\sqrt{2^{t}}}\sum_{j=0}^{2^{t}-1}\mathrm{e}^{2\pi \mathrm{i}j \frac{s}{r}} |j\rangle_{A}\right) (\mathbf{I}_B|u_s\rangle_{B})
=\frac{1}{\sqrt{r}}\sum_{s=0}^{r-1}|\psi_{s}\rangle_{A}|u_{s}\rangle_{B},
\end{align}
where $|\psi_{s}\rangle_A = \frac{1}{\sqrt{2^{t}}} \sum_{j=0}^{2^{t}-1} \mathrm{e}^{2\pi \mathrm{i} j \frac{s}{r}} (F^\dagger_A|j\rangle_A)=\frac{1}{2^{t}} \sum_{j=0}^{2^{t}-1}\sum_{k=0}^{2^{t}-1} \mathrm{e}^{2\pi \mathrm{i} j (\frac{s}{r}-\frac{k}{Q})} |k\rangle_A$ and $\mathbf{I}_B$ is the identity operator of register $B$. The measurement of the first register yields an approximation $\frac{k}{Q} \approx \frac{s}{r}$ where $s \in \{0, 1, \ldots, r-1\}$. Here, $|k\rangle_A$ represents a $t$-bit string that serves as an estimate of the fraction $\frac{s}{r}$ for some $s$. The probability of measuring outcome $k$ in the first register at step (iii) is given by
\begin{equation}\label{eq5}
P_k = \frac{1}{r} \sum_{s=0}^{r-1} \left| \frac{1}{Q}\sum_{j=0}^{2^{t}-1} \mathrm{e}^{2\pi \mathrm{i} j (\frac{s}{r}-\frac{k}{Q})} \right|^2.
\end{equation}
In essence, the order-finding algorithm randomly selects a value $s \in \{0, 1, \ldots, r-1\}$ and generates an approximation to $\frac{s}{r}$ expressed as $\frac{k}{Q}$. Subsequently, the order $r$ is determined by employing the continued fraction algorithm.


Denote by $\mathbb{R}^+$ the set of all nonnegative real numbers,
and let $f: \mathbb{R}^+ \to \mathbb{R}^+ $ be an operator monotone
function that satisfies $f(0)
> 0$ and $x f(1/x) = f(x)$. The corresponding Morozova-Chentsov
function is then defined as $c_f(x, y) = \frac{1}{y
f(x/y)}$\cite{PDM}. For any quantum state $\rho$ and operator $K$,
the generalization of metric-adjusted skew information is defined
by\cite{HFM}
\begin{equation}\label{eq6}
F_{f}(\rho,K)=\frac{f(0)}{2}\langle
\mathrm{i}[\rho,K],\mathrm{i}[\rho,K]\rangle_{\rho,f},
\end{equation}
where $\langle
A,B\rangle_{\rho,f}=\text{tr}[A^{\dagger}c_{f}(L_{\rho}R_{\rho})(B)]$
for any operators $A$ and $B$. Here $L_{\rho}(A)=\rho A$,
$R_{\rho}(A)=A\rho$, and $[A,B]=AB-BA$ is the commutator between
operators $A$ and $B$. For a pure state $\rho=|\psi\rangle
\langle\psi|$ and any operator $K$, $F_{f}(\rho,K)$ simply reduces
to
\begin{equation}\label{eq7}
I(|\psi\rangle\langle\psi|,K)=\frac{1}{2}[\langle\psi|K^{\dagger}K|\psi\rangle+\langle\psi|KK^{\dagger}|\psi\rangle]-|\langle\psi|K|\psi\rangle|^2,
\end{equation}
which reveals that $F_f(\rho, K)$ becomes independent of the
function $f$ for any pure state $\rho$, where $I(\rho, K):=
\frac{1}{2}\text{tr}[\sqrt{\rho},K][\sqrt{\rho},K]^\dag$ is the
modified Wigner-Yanase skew information\cite{LuoS2018,FanYJ,WuZ}.

A quantum channel is a trace preserving completely positive mapping,
which has the following Kraus representation
\begin{equation*}
\Phi(\rho) = \sum_{l=1}^m K_l \rho K_l^\dagger,
\end{equation*}
where the set $\{K_l\}_{l=1}^m$ comprises the Kraus operators
characterizing the channel, satisfying the completeness relation
$\sum_{l=1}^m K_l^\dagger K_l = \mathbf{I}$. A key focus in quantum
information theory is the quantification of coherence in relation to
quantum channels. Suppose that the density matrix $\rho$ has the
spectral decomposition $\rho = \sum_i \lambda_i
|\phi_i\rangle\langle\phi_i|$. The generalized metric-adjusted skew
information of $\rho$ with respect to a channel $\Phi$ is given by\cite{SYL}
\begin{equation}\label{eq8}
F_f(\rho, \Phi)= \sum_l F_f(\rho, K_l) = \frac{f(0)}{2} \sum_{ij} \frac{(\lambda_i - \lambda_j)^2}{\lambda_j f(\lambda_i/\lambda_j)} \langle\phi_i|\Phi(|\phi_j\rangle\langle\phi_j|)|\phi_i\rangle.
\end{equation}
When $\rho$ in Eq.~(\ref{eq8}) is a pure state, i.e., $\rho =
|\psi\rangle\langle\psi|$, $F_f(\rho,\Phi)$ reduces to
\begin{equation}\label{eq9}
F_f(|\psi\rangle\langle\psi|,\Phi)=\frac{1}{2}+\frac{1}{2}\sum_{l=1}^m
\langle\psi|K_lK_l^\dagger|\psi\rangle-\sum_{l=1}^m\left|\langle\psi|K_l|\psi\rangle\right|^2.
\end{equation}
In particular, when $f$ in Eq.~(\ref{eq8}) is chosen as the operator monotone function
\begin{equation*}
f(x) = \left(\frac{1 + \sqrt{x}}{2}\right)^2,
\end{equation*}
$F_f(\rho, \Phi)$ reduces to the
modified Wigner-Yanase skew information \cite{LuoS2018}
\begin{equation}\label{eq10}
I(\rho, \Phi) = \frac{1}{2}\sum_{l=1}^m \text{tr}\left[\sqrt{\rho},
K_l\right]\left[\sqrt{\rho},
K_l\right]^\dagger=\frac{1+\text{tr}(\rho\Phi(\mathbf{I}))}{2}-\text{tr}(\sqrt{\rho}\Phi(\sqrt{\rho})).
\end{equation}
Specifically, when $\rho$ in Eq.~(\ref{eq10}) is a pure state, i.e.,
$\rho = |\psi\rangle\langle\psi|$, one can find that
$I(|\psi\rangle\langle\psi|, \Phi)$ coincides with
$F_f(|\psi\rangle\langle\psi|,\Phi)$.

Rather than relying on a single quantity such as
the relative entropy of coherence, metric-adjusted skew information
offers a more flexible and fine-grained characterization by
exploiting the family structure induced by operator monotone
functions. Moreover, these quantities admit a clear physical
interpretation in parameter estimation\cite{HAS,BSL}, quantum
metrology\cite{GVS}, and asymmetry\cite{TRS}. Owing to their
intrinsic connection with quantum Fisher information, they can be
viewed not only as quantifiers of decoherence induced by quantum
channels, but also as measures directly linked to metrological
precision and sensitivity to parameter variations\cite{SYL}. In
contrast, the relative entropy of coherence can be interpreted as
the optimal rate for distilling a maximally coherent state from
given states\cite{WinterYang2016}.

In Shor's algorithm/generalized Shor's algorithm, we investigate the
coherence of final state $\rho'$ with respect to the von Neumann
measurement $\Pi = \{|i\rangle\langle i|_{AB}\}_{i=0}^{d-1}$
constructed from the orthonormal basis
$\{|i\rangle_{AB}\}_{i=0}^{d-1}$ of the $d$-dimensional Hilbert
space as
\begin{equation}\label{eq11}
C_f(\rho', \Pi) \triangleq F_f(\rho', \Pi).
\end{equation}
Correspondingly, the decoherence of the initial state $\rho$ induced by the
unitary transformation $\mathcal{S}$ is characterized by
\begin{equation}\label{eq12}
D_f(\rho,\mathcal{S}) \triangleq F_f(\rho,\mathcal{S}),
\end{equation}
where $\mathcal{S}(\cdot)=(F^\dagger_A\otimes \mathbf{I}_B)
U(\cdot)((F^\dagger_A\otimes \mathbf{I}_B)U)^{\dagger}$ represents the
composite unitary evolution conducted in Shor's algorithm. When
$\rho'$ and $\rho$ are pure states, the coherence of final state
$\rho'$ with respect to the von Neumann measurement is
\begin{equation}\label{eq13}
C(\rho', \Pi) \triangleq I(\rho', \Pi),
\end{equation}
and the decoherence of the initial state $\rho$ induced by the
unitary transformation $\mathcal{S}$ is
\begin{equation}\label{eq14}
D(\rho,\mathcal{S}) \triangleq I(\rho,\mathcal{S}).
\end{equation}

\vskip0.2in

\noindent {\bf 3. Coherence and decoherence in generalized Shor's algorithm}

In Section 3.1, we investigate the coherence and decoherence
dynamics and derive both lower and upper bounds on the performance
of Shor's algorithm when the register $A$ is initialized in an
arbitrary pure state. In Section 3.2, we investigate the coherence
and decoherence dynamics in Shor's algorithm when the register $AB$
is initialized in a pseudo-pure state. We also derive the
relationship between the probability of successfully calculating $r$
for an initial pure state in register $A$ and for an initial
pseudo-pure state in the combined register $AB$.

\vskip0.1in \noindent {\bf 3.1 Coherence and decoherence in Shor's algorithm for arbitrary pure state}\\\hspace*{\fill}\\
Instead of applying the Hadamard gate to obtain
$|\psi_{1}\rangle=\frac{1}{\sqrt{Q}}\sum_{j=0}^{Q-1}|j\rangle_A|1\rangle_B$,
we impose an arbitrary unitary quantum gate $U'$ on the initial
state $|0\rangle^{\otimes t}$ of register $A$, yielding the state in
the composite system $AB$ as
\begin{equation}\label{eq15}
|\phi_{1}\rangle = \sum_{j=0}^{Q-1} \alpha_j |j\rangle_A|1\rangle_B,
\end{equation}
where $\alpha_j$ are complex amplitudes satisfying $\sum_{j=0}^{Q-1}
|\alpha_j|^2 = 1$.

Applying the unitary transformation $U = \sum_{n=0}^{Q-1}
|n\rangle\langle n|_A \otimes U_B^n$, we then have
\begin{equation}\label{eq16}
|\phi_2\rangle = U\left(\sum_{j=0}^{Q-1} \alpha_j |j\rangle_A
|1\rangle_B\right) = \sum_{j=0}^{Q-1} \alpha_j |j\rangle_A |x^j \bmod
N\rangle_B.
\end{equation}
Noting that the eigenvectors of $U_{B}$ are $|u_{s}\rangle_B$,
$|\phi_2\rangle$ can be rewritten as
\begin{equation}\label{eq17}
U|\phi_{1}\rangle= \sum_{j=0}^{2^{t}-1}\alpha_j |j\rangle_A U_{B}^{j}\left(\frac{1}{\sqrt{r}} \sum_{s=0}^{r-1} |u_s\rangle_B\right) = \frac{1}{\sqrt{r}} \sum_{s=0}^{r-1} \sum_{j=0}^{2^{t}-1}\alpha_j\mathrm{e}^{2\pi \mathrm{i}j \frac{s}{r}} |j\rangle_A |u_s\rangle_B.
\end{equation}
Since $F^\dagger_A|j\rangle_A=\frac{1}{\sqrt{Q}}\sum_{k=0}^{Q-1}$ $
\mathrm{e}^{\frac{-2\pi \mathrm{i}jk}{Q}}$ $|k\rangle$, the state of
$AB$ after the inverse Fourier transform $F^\dagger_A$ is imposed on
register $A$ is
\begin{align}\label{eq18}
|\phi_3\rangle &= (F^\dagger_A \otimes \mathbf{I}_B)\left(\frac{1}{\sqrt{r}} \sum_{s=0}^{r-1} \sum_{j=0}^{2^{t}-1}\alpha_j\mathrm{e}^{2\pi \mathrm{i}j \frac{s}{r}} |j\rangle_A |u_s\rangle_B\right)\notag\\
&=\frac{1}{\sqrt{r}} \sum_{s=0}^{r-1} \sum_{j=0}^{2^{t}-1}\alpha_j\mathrm{e}^{2\pi \mathrm{i}j \frac{s}{r}}(F^\dagger_A |j\rangle_A ) (\mathbf{I}_B|u_s\rangle_B)\notag\\
&= \frac{1}{\sqrt{2^{t}r}} \sum_{s=0}^{r-1} \sum_{j,k=0}^{2^{t}-1}\alpha_j\mathrm{e}^{2\pi \mathrm{i}j (\frac{s}{r}-\frac{k}{Q})} |k\rangle_A|u_s\rangle_B.
\end{align}
Note that by applying the inverse Fourier transform $F^\dagger_A$ to
Eq. (\ref{eq16}), we also obtain
\begin{align}\label{eq19}
|\phi_3\rangle &= (F^\dagger_A \otimes \mathbf{I}_B)\left(\sum_{j=0}^{Q-1} \alpha_j |j\rangle_A |x^j \bmod N\rangle_B\right)\notag\\
&= \frac{1}{\sqrt{Q}}\sum_{k=0}^{Q-1}\sum_{j=0}^{Q-1} \alpha_j
\mathrm{e}^{-2\pi \mathrm{i}jk/Q}|k\rangle_A |x^j \bmod N\rangle_B.
\end{align}

When measuring register $A$, the probability of obtaining outcome
$k$ is
\begin{equation}\label{eq20}
P(k) = \frac{1}{Qr}\sum_{s=0}^{r-1}
\left|\sum_{j=0}^{2^{t}-1}\alpha_j \mathrm{e}^{2\pi \mathrm{i}j(s/r -
k/Q)}\right|^2.
\end{equation}
The quantum circuit for Shor's algorithm in this scenario is
illustrated in Figure 1.

\begin{figure}[H]\centering
{\begin{minipage}[figure1]{0.7\linewidth}
\hspace{-0.1\textwidth}
\includegraphics[width=1.1\textwidth,natwidth=12cm,natheight=6cm]{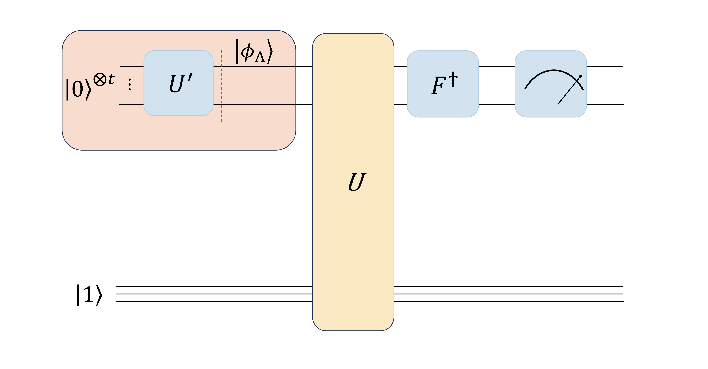}
\end{minipage}}
\caption{ Quantum circuit for Shor's algorithm when
the register $A$ is initialized in an arbitrary pure state. \label{fig:Fig1}}
\end{figure}
To simplify the subsequent analysis, we assume that $Q = rm$, where
$m$ is a positive integer. Then it follows that
$\left\lfloor\frac{Q-1}{r}\right\rfloor = m-1$. By decomposing the
index $j$ in the form $j = a + br$, where $a$ and $b$ serve as the
remainder and quotient respectively, we can reformulate the quantum
state $|\phi_3\rangle$, that is,
\begin{equation*}
|\phi_3\rangle = \frac{1}{\sqrt{Q}} \sum_{k=0}^{Q-1}
\sum_{a=0}^{r-1} \sum_{b=0}^{m-1} \alpha_{a+br}\mathrm{e}^{-2\pi
\mathrm{i}(a+br)k/Q} |k\rangle_A|x^a\bmod N\rangle_B,
\end{equation*}
where $0\leq a\leq r-1$ and $0\leq b\leq m-1$. The probability of
obtaining the measurement result $k$ can be rewritten as
\begin{equation}\label{eq21}
P(k)=\frac{1}{Q}\sum_{a=0}^{r-1} \left|\sum_{b=0}^{m-1} \alpha_{a+br}\mathrm{e}^{-2\pi \mathrm{i}brk/Q}\right|^2.
\end{equation}
Denote $A_{a,k}=\sum_{b=0}^{m-1} \alpha_{a+br}\mathrm{e}^{-2\pi \mathrm{i}brk/Q}$. Then we have $P(k)=\frac{1}{Q}\sum_{a=0}^{r-1} \left|A_{a,k}\right|^2.$\\
{\bf Remark 1} When $k = s \cdot\frac{Q}{r}$ $=sm$, $s=\{0,1,\cdots,
r-1\}$, we have $\mathrm{e}^{-2\pi \mathrm{i}brk/Q} = \mathrm{e}^{-2\pi
\mathrm{i}bs} = 1$, and the formula simplifies to
\begin{equation}\label{eq22}
P(k) = \frac{1}{Q}\sum_{a=0}^{r-1} \left|\sum_{b=0}^{m-1} \alpha_{a+br}\right|^2.
\end{equation}

We now proceed to investigate the coherence and decoherence in generalized Shor's algorithm when the register $A$ is initialized in an arbitrary pure state. \\
{\bf Theorem 1} For the generalized Shor's algorithm with the
register $A$ initialized in an arbitrary pure state, the coherence
of state $|\phi_{3}\rangle$ with respect to the von Neumann
measurement $\Pi$ in the computational basis is quantified by
\begin{equation}\label{eq23}
C(\rho'_{3}, \Pi)
=1-\sum_{a=0}^{r-1}\sum_{k=0}^{Q-1} \frac{\left|A_{a,k}\right|^4}{Q^{2}},
\end{equation}
where $\Pi(\cdot)=\sum_{i} K_{i}(\cdot) K_{i}^{\dagger}$,
$K_{i}=|i\rangle \langle i|_{AB}$, $\rho'_{3}$ represents the density
operator of $|\phi_{3}\rangle$ and $A_{a,k}=\sum_{b=0}^{m-1}
\alpha_{a+br}\mathrm{e}^{-2\pi \mathrm{i}brk/Q}$, and the decoherence of the
initial state $\rho'_{1}$ induced by the unitary transformation
$\mathcal{S}$ is characterized by
\begin{equation}\label{eq24}
D(\rho'_{1},\mathcal{S})
=1-\frac{1}{Q}\left|\sum_{k=0}^{Q-1}\alpha_{k}A_{0,k}\right|^2,
\end{equation}
where $\rho'_{1}$ represents the density operator of $|\phi_{1}\rangle$ and $A_{0,k}=\sum_{b=0}^{m-1} \alpha_{br}\mathrm{e}^{-2\pi \mathrm{i}brk/Q}$.\\
$\it{Proof}$. Let the density matrix of $|\phi_3\rangle$ be
$\rho'_{3}$. According to Eqs. (\ref{eq9}), (\ref{eq13}) and
(\ref{eq19}), by letting $A_{a,k}=\sum_{b=0}^{m-1}
\alpha_{a+br}\mathrm{e}^{-2\pi \mathrm{i}brk/Q}$, we have
\begin{align*}
C(\rho'_{3}, \Pi)
&=1-\sum_{a=0}^{r-1}\sum_{k=0}^{Q-1} \left|\frac{1}{Q}\sum_{b=0}^{m-1} \sum_{b^{'}=0}^{m-1} \alpha_{a+br}\alpha_{a+b^{'}r}\mathrm{e}^{-2\pi \mathrm{i}(b-b^{'})rk/Q}\right|^2\notag\\
&=1-\sum_{a=0}^{r-1}\sum_{k=0}^{Q-1} \frac{\left|A_{a,k}\right|^4}{Q^{2}}.
\end{align*}
Let the density matrix of $|\phi_1\rangle$ be $\rho'_{1}$. According
to Eqs. (\ref{eq9}), (\ref{eq14}) and (\ref{eq15}), by letting
$A_{0,k}=\sum_{b=0}^{m-1} \alpha_{br}$ $\mathrm{e}^{-2\pi
\mathrm{i}brk/Q}$, we have
\begin{align*}
D(\rho'_{1},\mathcal{S})
&=1-\frac{1}{Q}\left|\sum_{k=0}^{Q-1}\sum_{b=0}^{m-1} \alpha_{k}\alpha_{br}\mathrm{e}^{-2\pi \mathrm{i}brk/Q}\right|^2\notag\\
&=1-\frac{1}{Q}\left|\sum_{k=0}^{Q-1}\alpha_{k}A_{0,k}\right|^2.
\end{align*}
\qed

Theorem 1 characterizes coherence and decoherence of arbitrary pure initial states in quantum systems. It precisely quantifies the coherence of state $\rho'_{3}$ under von Neumann measurements and the decoherence of quantum operation $S(\cdot)$ on initial state $\rho'_1$.\\
\noindent{\bf Theorem 2} For the generalized Shor's algorithm with the register $A$ initialized in an arbitrary pure state, the probability that we can calculate $r$ satisfies
\begin{equation}\label{eq25}
P> \frac{4Q|\alpha_{\min}|^{2}\phi(r)}{r\pi^2},
\end{equation}
where $|\alpha_{\min}|^{2}=\min_{j}|\alpha_{j}|^{2}$, $\phi$ is Euler's totient function, and $\phi(r)$ is the number of integers $s$ that are less than and coprime to $r$.\\
$\it{Proof}$. According to Eq. (\ref{eq20}), the probability of
obtaining outcome $k$ in register $A$ satisfies
\begin{equation*}
P(k)>\frac{|\alpha_{\min}|^{2}}{Qr}\sum_{s=0}^{r-1}
\left|\sum_{j=0}^{2^{t}-1}\mathrm{e}^{2\pi \mathrm{i}j(s/r - k/Q)}\right|^2,
\end{equation*}
where $|\alpha_{\min}|^{2}=\min_{j}|\alpha_{j}|^{2}$. Denote
$\sum_{j=0}^{2^{t}-1} \mathrm{e}^{2\pi \mathrm{i}j(s/r - k/Q)}$ by
$S$. Since $S$ represents an arithmetic progression, the magnitude
squared $|S|^2$ can be effectively bounded as
\begin{equation*}
|S|^2 > \frac{4Q^2}{\pi^2} \quad \text{for} \quad \left|\frac{s}{r}-\frac{k}{Q}\right| < \frac{1}{2Q}.
\end{equation*}
Since the condition $Q > N^2$ holds, this constitutes a sufficient condition ensuring that, given the ratio $\frac{k}{Q}$, there exists exactly one fraction $\frac{s}{r}$ with $r < N$ that satisfies the aforementioned condition.
Thus the probability that $\frac{k}{Q}$ is the best estimate of a fraction with
denominator $r$ is
\begin{equation*}
P'(k)>\frac{|\alpha_{\min}|^{2}}{Qr}\cdot\frac{4Q^2}{\pi^2}
=\frac{4Q|\alpha_{\min}|^{2}}{r\pi^2}.
\end{equation*}
For the algorithm to function correctly, we impose the constraint
that the numerator $s$ is coprime to $r$. This requirement is
essential because, in the absence of this condition, cancellation of
common factors would occur in the fraction $s/r$, thereby
compromising the algorithm's effectiveness.

The number of values of $s$ that satisfy both conditions---being less than $r$ and coprime to $r$---is given by Euler's totient function $\phi(r)$\cite{Hardy}. Consequently, the probability of successfully calculating $r$ can be estimated by Eq. (\ref{eq25}).\qed

Specifically, Hardy and Wright's theorem  provides the asymptotic
bound $\frac{\phi(r)}{r} > \frac{\delta}{\log \log r}$ \cite{Hardy},
where $\delta$ is a positive constant. So by Theorem 2 we can
further derive that $P> \frac{4Q|\alpha_{\min}|^{2}\delta}{\pi^2\log
\log r}$.

To give the upper bound of the probability that we can calculate
$r$, we first give an estimation of the probability of obtaining
outcome $k$.

\noindent {\bf Lemma 1} The probability of obtaining outcome $k$ in
register $A$ satisfies the following relation
\begin{equation}\label{eq26}
1 - C(\rho'_3, \Pi)\leq\sum_{k} P(k)^2 \leq r( 1 - C(\rho'_3,
\Pi)),
\end{equation}
where the upper bound is saturated when the initial state of
register $A$ is $|\phi\rangle_A=|+\rangle^{\otimes t}_A =
\sum_{j=0}^{Q-1}\frac{1}{\sqrt{Q}} |j\rangle_A$, and
the lower bound is saturated when the initial state of register $A$ is $|\phi\rangle_A = |j\rangle_A$.\\
$\it{Proof}$. From Theorem 1 and Eq. (\ref{eq21}), we have
\begin{equation*}
\sum_{a=0}^{r-1} |A_{a,k}|^2 =  QP(k)~~~ \mathrm{and} ~~~
\sum_{a=0}^{r-1} \sum_{k=0}^{Q-1} |A_{a,k}|^4=Q^2( 1 - C(\rho'_3, \Pi) ) .
\end{equation*}
By applying the Cauchy-Schwartz inequality for the sequence $ \{ |A_{a,k}| \} $, we have
\begin{equation*}
\left( \sum_{a=0}^{r-1} |A_{a,k}|^{2} \right)^2 \leq r \sum_{a=0}^{r-1} |A_{a,k}|^4.
\end{equation*}
Since $\left( \sum_{a=0}^{r-1} |A_{a,k}|^{2} \right)^2=Q^2 P(k)^2$,
we obtain
\begin{equation*}
Q^2 P(k)^2 \leq r \sum_{a=0}^{r-1} |A_{a,k}|^4.
\end{equation*}
Summing over $k$ yields
\begin{equation*}
Q^2 \sum_{k=0}^{Q-1} P(k)^2 \leq r \sum_{k=0}^{Q-1} \sum_{a=0}^{r-1} |A_{a,k}|^4 = r Q^2 \left( 1 - C(\rho'_3, \Pi) \right).
\end{equation*}
which gives an upper bound for $\sum_{k} P(k)^2$ as
\begin{equation*}
\sum_{k} P(k)^2 \leq r( 1 - C(\rho'_3, \Pi)).
\end{equation*}
We next prove the corresponding lower bound. From Eq. (\ref{eq21}),
we can obtain
\begin{equation*}
P(k)^2 =\frac{1}{Q^2}\left(\sum_a |A_{a,k}|^2\right)^2.
\end{equation*}
Expanding the squared term yields
\begin{equation*}
\left(\sum_a |A_{a,k}|^2\right)^2=\sum_a\sum_{a'} |A_{a,k}|^2 |A_{a',k}|^2= \sum_a |A_{a,k}|^4 + \sum_{a \neq a'} |A_{a,k}|^2 |A_{a',k}|^2.
\end{equation*}
Substituting this expansion into the expression for $P(k)^2$ and summing over all $k$, we obtain
\begin{equation*}
\sum_k P(k)^2 = \frac{1}{Q^2} \sum_k \left(\sum_a |A_{a,k}|^4 + \sum_{a \neq a'} |A_{a,k}|^2 |A_{a',k}|^2 \right).
\end{equation*}
Noting that
\begin{equation*}
1 -C(\rho'_3, \Pi)=\frac{1}{Q^2}\sum_{k}\sum_{a} |A_{a,k}|^4,
\end{equation*}
we get
\begin{equation*}
\sum_k P(k)^2 \geq 1 - C(\rho'_3, \Pi).
\end{equation*}
Thus we have derived Eq. (\ref{eq26}).

Note that $\sum_{k} P(k)^2 = r( 1 - C(\rho'_3, \Pi))$ if and only if
for any given $k$, $|A_{a,k}|=\alpha$ for all $a\in \{0, 1, \ldots,
r-1\}$, where $\alpha$ is a constant. It is easy to check that the
upper bound is saturated when the initial state of register $A$ is
$|\phi\rangle_A= |+\rangle^{\otimes
t}_A=\sum_{j=0}^{Q-1}\frac{1}{\sqrt{Q}} |j\rangle_A $.

On the other hand, $\sum_{k} P(k)^2 = 1-C(\rho'_3, \Pi)$ if and only if
for each $k$, $|A_{a,k}|=0$ or $|A_{a',k}|=0$ for $a\neq a'$. It is easy to check that the lower bound is saturated when the initial state of register $A$ is $|\phi\rangle_A = |j\rangle_A$.\qed\\
{\bf Remark 2} From Lemma 1, the lower and upper bounds of the sum of $P(k)^2$ are closely related to the coherence $C(\rho'_3, \Pi)$, indicating that $\sum_{k} P(k)^2$ cannot be too large if the coherence $C(\rho'_3, \Pi)$ is large and it cannot be arbitrarily small when the coherence $C(\rho'_3, \Pi)$ is very small.\\
{\bf Theorem 3} For the generalized Shor's algorithm with the
register $A$ initialized in an arbitrary pure state, the
probability of calculating $r$ satisfies the following relation
\begin{equation}\label{eq27}
P^{2}<r(1 - C(\rho'_3, \Pi))\phi^{2}(r).
\end{equation}
$\it{Proof}$. The probability that $\frac{k}{Q}$ is the best estimate of a fraction with
denominator $r$, i.e., $\left|\frac{s}{r}-\frac{k}{Q}\right| < \frac{1}{2Q}$, satisfies
\begin{equation*}
P'^{2}(k)<\sum_{k} P(k)^2.
\end{equation*}
We now require that the numerator, $s$, be coprime to $r$.
Then, the probability of the generalized Shor's algorithm that we can calculate $r$ is
\begin{equation*}
P^{2}<\phi^{2}(r)\sum_{k} P(k)^2.
\end{equation*}
According to Lemma 1, $ \sum_{k}P(k)^2 \leq r( 1 - C(\rho'_3,
\Pi))$, and thus we get Eq. (\ref{eq27}). \qed

We now consider a special case of the global unitary operation $U'$,
that is, the local unitary operation acting on an $t$-qubit quantum
register as
\begin{equation*}
U' = \left(U'(\alpha, \beta, \theta)\right)^{\otimes t}.
\end{equation*}
Here, the single-qubit unitary gate $U'(\alpha, \beta, \theta)$ is
parameterized by three real parameters $\alpha$, $\beta$, and
$\theta$, where $\alpha, \beta, \theta \in [0, \pi/2]$, which is of
the form
\begin{equation*}
U'(\alpha, \beta, \theta) = \left[\begin{matrix}
\mathrm{e}^{\mathrm{i}\alpha} \cos \theta & \mathrm{e}^{-\mathrm{i}\beta} \sin \theta \\
\mathrm{e}^{\mathrm{i}\beta} \sin \theta & -\mathrm{e}^{-\mathrm{i}\alpha} \cos \theta
\end{matrix}
\right].
\end{equation*}
It is worth noting that when $\alpha = \beta = 0$ and $\theta = \pi/4$, this gate reduces to the well-known Hadamard operator.
When we apply this unitary gate to the single qubit state $|0\rangle$, the resulting state is
\begin{equation*}
|\phi(\alpha, \beta, \theta)\rangle = U'(\alpha, \beta, \theta)|0\rangle
= \mathrm{e}^{\mathrm{i}\alpha} \cos \theta |0\rangle + \mathrm{e}^{\mathrm{i}\beta} \sin \theta |1\rangle.
\end{equation*}
In this case, the initial state and the final state of the register
$AB$ is
\begin{equation*}
|\phi_{1}\rangle = |\phi(\alpha, \beta, \theta)\rangle^{\otimes
t}_A|1\rangle_B = \sum_{j=0}^{Q-1} \left(\mathrm{e}^{\mathrm{i}\alpha} \cos
\theta\right)^{x_j} \left(\mathrm{e}^{\mathrm{i}\beta} \sin
\theta\right)^{t-x_j} |j\rangle_A|1\rangle_B
\end{equation*}
and
\begin{equation*}
|\phi_3\rangle=\frac{1}{\sqrt{Q}}\sum_{k=0}^{Q-1}\sum_{j=0}^{Q-1}
\left(\mathrm{e}^{\mathrm{i}\alpha} \cos \theta\right)^{x_j}
\left(\mathrm{e}^{\mathrm{i}\beta} \sin \theta\right)^{t-x_j}
\mathrm{e}^{-2\pi \mathrm{i}jk/Q}|k\rangle_A |x^j \bmod N\rangle_B,
\end{equation*}
respectively, where $x_j$ denotes the number of zeros in the binary
representation of $j = j_1 j_2 \cdots j_t$. Moreover, the coherence
of state $|\phi_{3}\rangle$ with respect to the von Neumann
measurement $\Pi$ is
\begin{equation}\label{eq28}
C(\rho'_{3}, \Pi)
=1-\sum_{a=0}^{r-1}\sum_{k=0}^{Q-1} \frac{\left|E_{a,k}\right|^4}{Q^{2}},
\end{equation}
where $E_{a,k}=\sum_{b=0}^{m-1}\left(\mathrm{e}^{\mathrm{i}\alpha} \cos
\theta\right)^{x_{a+br}} \left(\mathrm{e}^{\mathrm{i}\beta} \sin
\theta\right)^{t-x_{a+br}}\mathrm{e}^{-2\pi \mathrm{i}brk/Q}$, and the
decoherence of the initial state $\rho'_{1}$ induced by
$\mathcal{S}$ is
\begin{equation}\label{eq29}
D(\rho'_{1},\mathcal{S})
=1-\frac{1}{Q}\left|\sum_{k=0}^{Q-1}\left(\mathrm{e}^{\mathrm{i}\alpha} \cos \theta\right)^{x_k} \left(\mathrm{e}^{\mathrm{i}\beta} \sin \theta\right)^{t-x_k}G_{0,k}\right|^2,
\end{equation}
where $G_{0,k}=\sum_{b=0}^{m-1}\left(\mathrm{e}^{\mathrm{i}\alpha} \cos \theta\right)^{x_br} \left(\mathrm{e}^{\mathrm{i}\beta} \sin \theta\right)^{t-x_br}\mathrm{e}^{-2\pi \mathrm{i}brk/Q}$.

When $\theta < \pi/4$, we have
$|\alpha_{\min}|^{2}=\min_{j}|\alpha_{j}|^{2}=\sin^{2t} \theta$, and
the probability that we can calculate $r$ is $P> \frac{4Q\sin^{2t}
\theta\phi(r)}{r\pi^2}$. When $\theta > \pi/4$, we have
$|\alpha_{\min}|^{2}=\min_{j}|\alpha_{j}|^{2}=\cos^{2t} \theta$, and
the probability that we can calculate $r$ is $P> \frac{4Q\cos^{2t}
\theta\phi(r)}{r\pi^2}$. When $\theta = \pi/4$, we have
$|\alpha_{\min}|^{2}=\min_{j}|\alpha_{j}|^{2}=\frac{1}{Q}$, and the
probability that we can calculate $r$ is $P>
\frac{4\phi(r)}{r\pi^2}$.

Note that the state $|\phi_1\rangle$ here corresponds to the state
in Eq. (\ref{eq15}) in which $\alpha_j=\left(\mathrm{e}^{\mathrm{i}\alpha}
\cos \theta\right)^{x_j}
\left(\mathrm{e}^{\mathrm{i}\beta} \sin \theta\right)^{t-x_j}$. So when $\alpha = \beta = 0$ and $\theta = \frac{\pi}{4}$, $U'(\alpha, \beta, \theta)$ reduces to the Hadamard operator, which corresponds to the original Shor's algorithm where all $\alpha_j = \frac{1}{\sqrt{Q}}$. This leads to the following result.\\
{\bf Corollary 1} The coherence of state $|\psi_{3}\rangle$ with respect to the von Neumann measurement $\Pi$ in the computational basis is given by
\begin{equation}\label{eq30}
C(\rho_{3}, \Pi)=1-\frac{1}{r^{2}},
\end{equation}
where $\rho_{3}$ is the density operator of $|\psi_{3}\rangle$ and
$r$ is the order of Shor's algorithm, and the decoherence of the initial state $\rho'_{1}$ induced by $\mathcal{S}$ is
\begin{equation}\label{eq31}
D(\rho_{1},\mathcal{S})=1-\frac{1}{Q},
\end{equation}
where $\rho_{1}$ is the density operator of $|\psi_{1}\rangle$ and
$Q = 2^t$ is the dimension of register $A$.

\vskip0.1in \noindent {\bf 3.2 Coherence and decoherence in Shor's algorithm for any pseudo-pure state}\\\hspace*{\fill}\\
Pseudo-pure states can be effectively prepared using existing
techniques\cite{Linden2001,Cory1997,Sharf2000}, and play a crucial
role in quantum computation in the presence of noise. Indeed, the
utility of pseudo-pure states for nuclear magnetic resonance (NMR)
quantum computation has been recognized \cite{Braunstein1999}. It is
shown that an entanglement-free version of the Bernstein-Vazirani
(BV) algorithm on an NMR ensemble quantum computer can be
implemented\cite{Du2001}, and for a given bounded purity,
pseudo-pure states are the quantum states that achieve optimal
performance in the probabilistic BV algorithm \cite{NMTV}.

Consider Shor's algorithm with the register $AB$ initialized in a
pseudo-pure state\cite{Cory1997}
\begin{equation}\label{eq32}
\rho^{\varepsilon}_1 = (1 - \varepsilon)|\phi_1\rangle\langle\phi_1| + \frac{\varepsilon}{d}\mathbf{I}_{AB},
\end{equation}
where $0\leq\varepsilon\leq1$, $|\phi_1\rangle = \sum_{j=0}^{Q-1}
\alpha_j|j\rangle_A|1\rangle_B$ and $d=2^n$ denotes the dimension of the
composite system $AB$. It is one of the simplest models for
mixed-state quantum computation. After applying the unitary
transformation $U = \sum_{n=0}^{Q-1} |n\rangle\langle n|_A \otimes
U_B^n$, the resulting state becomes
\begin{equation}\label{eq33}
\rho^{\varepsilon}_2 = (1 - \varepsilon)|\phi_2\rangle\langle\phi_2| + \frac{\varepsilon}{d}\mathbf{I}_{AB},
\end{equation}
where $|\phi_2\rangle = \sum_{j=0}^{Q-1} \alpha_j|j\rangle_A|x^j \bmod
N\rangle_B$. Applying the inverse Fourier transform $F^{\dagger}$ to
register $A$ yields
\begin{equation}\label{eq34}
\rho^{\varepsilon}_3 = (1 - \varepsilon)|\phi_3\rangle\langle\phi_3| + \frac{\varepsilon}{d}\mathbf{I}_{AB},
\end{equation}
where$|\phi_3\rangle = \frac{1}{\sqrt{Q}} \sum_{k=0}^{Q-1} \sum_{j=0}^{Q-1} \alpha_j \mathrm{e}^{-2\pi \mathrm{i}jk/Q}|k\rangle_A|x^j \bmod N\rangle_B$. When measuring register $A$, the probability of obtaining outcome $k$ is
\begin{equation}\label{eq35}
P_{\varepsilon}(k) = (1 - \varepsilon)P(k) + \frac{\varepsilon}{Q}.
\end{equation}
\begin{figure}[H]\centering
\hspace{-0.15\textwidth}
{\begin{minipage}[figure2]{0.8\linewidth}
\includegraphics[width=1.1\textwidth,natwidth=12cm,natheight=6cm]{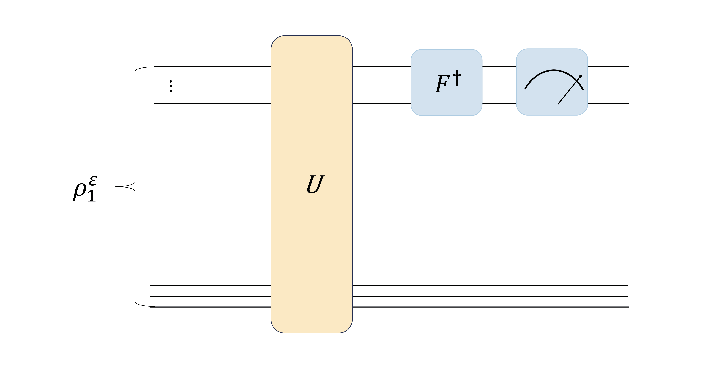}
\end{minipage}}
\caption{ Quantum circuit for Shor's algorithm when
the register $AB$ is initialized in a pseudo-pure state. \label{fig:Fig1}}
\end{figure}
For large $Q$, the probability of obtaining outcome $k$ with the register $AB$ initialized in a pseudo-pure state is approximately proportional to the probability of obtaining outcome $k$ with the register $A$ initialized in an arbitrary pure state.
The quantum circuit for Shor's algorithm in this scenario is illustrated in Figure 2.\\
{\bf Theorem 4} For the generalized Shor's algorithm with the
register $AB$ initialized in a pseudo-pure state, the coherence of
state $\rho^{\varepsilon}_3$ with respect to the von Neumann
measurement $\Pi$ in the computational basis is quantified by
\begin{equation}\label{eq36}
C_{f}(\rho^{\varepsilon}_3, \Pi)= \frac{f(0)(1-\varepsilon)^2C(\rho'_{3}, \Pi)}{2} \left[ \frac{d}{\varepsilon \cdot f\left(\frac{d-(d-1)\varepsilon}{\varepsilon}\right)} + \frac{d}{(d-(d-1)\varepsilon) \cdot f\left(\frac{\varepsilon}{d-(d-1)\varepsilon}\right)} \right],
\end{equation}
where $0< \varepsilon\leq1$ and $C(\rho'_{3}, \Pi)$ represents the
coherence of $|\phi_{3}\rangle$, and the decoherence of the initial
state $\rho^{\varepsilon}_1$ induced by the unitary transformation
$\mathcal{S}$ is
\begin{equation}\label{eq37}
D_{f}(\rho^{\varepsilon}_1,\mathcal{S})
= \frac{f(0)(1-\varepsilon)^2D(\rho'_{1}, \Pi)}{2} \left[ \frac{d}{\varepsilon \cdot f\left(\frac{d-(d-1)\varepsilon}{\varepsilon}\right)} + \frac{d}{(d-(d-1)\varepsilon) \cdot f\left(\frac{\varepsilon}{d-(d-1)\varepsilon}\right)} \right],
\end{equation}
where $0< \varepsilon\leq1$ and $D(\rho'_{1}, \Pi)$ represents the decoherence of $|\phi_{1}\rangle$.\\
$\it{Proof}$. The spectral decomposition of the state $\rho^{\varepsilon}_3$ takes the form
\begin{equation*}
\rho^{\varepsilon}_3 = \lambda_1|\phi_3\rangle\langle\phi_3| + \sum_{j=2}^d \lambda_j|\varphi_j\rangle\langle\varphi_j|,
\end{equation*}
where $\lambda_1= 1 - \frac{(d-1)\varepsilon}{d}$ and $\lambda_j =
\frac{\varepsilon}{d}$ $(j\in \{2, 3, \ldots, d\})$. Here,
$\{|\varphi_j\rangle\}_{j=1}^d$ represents the orthonormal basis,
where $|\varphi_1\rangle=|\phi_3\rangle$ and the remaining states
$|\varphi_j\rangle$ satisfy $\langle\varphi_j|\phi_3\rangle = 0$ for
$j \geq 2$. Utilizing the orthogonality relation $\sum_{j=2}^d
|\varphi_j\rangle\langle\varphi_j| = \mathbf{I}_{AB} -
|\phi_3\rangle\langle\phi_3|$, we can rewrite this as
\begin{equation*}
\rho^{\varepsilon}_3 = \left(1 - \frac{(d-1)\varepsilon}{d}\right)|\phi_3\rangle\langle\phi_3| + \frac{\varepsilon}{d}\left(\mathbf{I}_{AB} - |\phi_3\rangle\langle\phi_3|\right).
\end{equation*}
According to Eqs. (\ref{eq8}) and (\ref{eq11}), the coherence of state $\rho^{\varepsilon}_3$ with respect to the von Neumann measurement $\Pi$ is
\begin{equation*}
C_{f}(\rho^{\varepsilon}_3, \Pi)= \sum_i F_f(\rho, K_i) = \frac{f(0)}{2} \sum_{jk} \frac{(\lambda_j - \lambda_k)^2}{\lambda_k f(\lambda_j/\lambda_k)} \langle\varphi_j|\Pi(|\varphi_k\rangle\langle\varphi_k|)|\varphi_j\rangle.
\end{equation*}
After algebraic manipulation, the result of $C_{f}(\rho^{\varepsilon}_3, \Pi)$ is
\begin{equation*}
\frac{f(0)(1-\varepsilon)^2}{2} \left[ \frac{d}{\varepsilon \cdot
f\left(\frac{d-(d-1)\varepsilon}{\varepsilon}\right)} +
\frac{d}{(d-(d-1)\varepsilon) \cdot
f\left(\frac{\varepsilon}{d-(d-1)\varepsilon}\right)} \right]\left(
1 - \sum_{i=0}^{d-1} |\langle i|\phi_3\rangle|^4 \right),
\end{equation*}
where $0< \varepsilon\leq1$, and $\langle i|$ denotes $\langle
i|_{AB}$ for the sake of simplicity. The coherence of
$|\phi_{3}\rangle$ with respect to the von Neumann measurement $\Pi$
is $1 - \sum_{i=0}^{d-1} |\langle i|\phi_3\rangle|^4$. Thus we can
drive Eq. (\ref{eq36}).
 The spectral decomposition of the state $\rho^{\varepsilon}_1$ takes the form
\begin{equation*}
\rho^{\varepsilon}_1 = \lambda_1|\phi_1\rangle\langle\phi_1| + \sum_{j=2}^d \lambda_j|\varphi'_j\rangle\langle\varphi'_j|,
\end{equation*}
where $\lambda_1= 1 - \frac{(d-1)\varepsilon}{d}$ and $\lambda_j = \frac{\varepsilon}{d}$ $(j\in \{2, 3, \ldots, d\})$. Here, $\{|\varphi'_j\rangle\}_{j=1}^d$ is an orthonormal basis with $|\varphi'_1\rangle=|\phi_1\rangle$ and $\langle\varphi'_j|\phi_1\rangle = 0$ for $j \geq 2$.
According to Eqs. (\ref{eq8}) and (\ref{eq12}), the decoherence of the initial state $\rho^{\varepsilon}_1$ induced by the unitary transformation $\mathcal{S}$ is
\begin{equation*}
D_{f}(\rho^{\varepsilon}_1, S)= F_f(\rho^{\varepsilon}_1, S) = \frac{f(0)}{2} \sum_{jk} \frac{(\lambda_j - \lambda_k)^2}{\lambda_k f(\lambda_j/\lambda_k)} \langle\varphi'_j|S(|\varphi'_k\rangle\langle\varphi'_k|)|\varphi'_j\rangle.
\end{equation*}
The value of decoherence $D_{f}(\rho^{\varepsilon}_1, S)$ is
\begin{equation*}
\frac{f(0)(1-\varepsilon)^2}{2} \left[ \frac{d}{\varepsilon \cdot f\left(\frac{d-(d-1)\varepsilon}{\varepsilon}\right)} + \frac{d}{(d-(d-1)\varepsilon) \cdot f\left(\frac{\varepsilon}{d-(d-1)\varepsilon}\right)} \right]\left( 1 - |\langle \phi_1|\phi_3\rangle|^2 \right),
\end{equation*}
where $0< \varepsilon\leq1$.
Noting that the decoherence of the initial state $\rho'_{1}$ induced by the unitary transformation $\mathcal{S}$ is $1-|\langle \phi_1|\phi_3\rangle|^2$, we obtain Eq. (\ref{eq37}). \qed

After algebraic manipulation, the square root is given by
\begin{equation*}
\sqrt{\rho^{\varepsilon}_3} = \sqrt{\frac{\varepsilon}{d}}\mathbf{I}_{AB} + \left(\sqrt{1 - \frac{(d-1)\varepsilon}{d}} - \sqrt{\frac{\varepsilon}{d}}\right)|\phi_3\rangle\langle\phi_3|.
\end{equation*}
Let the operator monotone function be $f(x) = \left(\frac{1 + \sqrt{x}}{2}\right)^2$. Then we can obtain the following result.\\
{\bf Corollary 2} The coherence of state $\rho^{\varepsilon}_3$ with respect to the von Neumann measurement $\Pi$ is given by
\begin{equation}\label{eq38}
C(\rho^{\varepsilon}_3, \Pi)= \left(\sqrt{1 - \frac{(d-1)\varepsilon}{d}} - \sqrt{\frac{\varepsilon}{d}}\right)^2C(\rho'_{3}, \Pi),
\end{equation}
where $C(\rho'_{3}, \Pi)$ represents the coherence of
$|\phi_{3}\rangle$, and the decoherence of the initial state
$\rho^{\varepsilon}_1$ induced by the unitary transformation
$\mathcal{S}$ is
\begin{equation}\label{eq39}
D(\rho^{\varepsilon}_1,\mathcal{S})
= \left(\sqrt{1 - \frac{(d-1)\varepsilon}{d}} - \sqrt{\frac{\varepsilon}{d}}\right)^{2}D(\rho'_{1}, S),
\end{equation}
where $D(\rho'_{1}, S)$ represents the decoherence of $|\phi_{1}\rangle$. \\
{\bf Theorem 5} The probability that we can calculate $r$ when the
register $AB$ is initialized in a pseudo-pure state and the one when
the register $A$ is initialized in an arbitrary pure state exhibit
the following relation
\begin{equation}\label{eq40}
P(\rho^{\varepsilon}_3)= (1 - \varepsilon)P + \frac{\varepsilon\phi(r)}{Q}.
\end{equation}
$\it{Proof}$. According to Eq. (\ref{eq35}), the probability that $\frac{k}{Q}$ is the best estimate of a fraction with
denominator $r$, i.e., $\left|\frac{s}{r}-\frac{k}{Q}\right| < \frac{1}{2Q}$, satisfies
\begin{equation*}
P'_{\varepsilon}(k)=(1 - \varepsilon)P'(k) + \frac{\varepsilon}{Q}.
\end{equation*}
We now require that the numerator $s$ is coprime to $r$. There are
$\phi(r)$ values of $s$ which are less than and coprime to $r$.
Then, the probability that we can calculate $r$ when the register $AB$ is initialized in a pseudo-pure state is
\begin{equation*}
P(\rho^{\varepsilon}_3)=P'_{\varepsilon}(k)\phi(r)= (1 - \varepsilon)P'(k) \phi(r)+ \frac{\varepsilon\phi(r)}{Q}.
\end{equation*}
Since $ P'(k) \phi(r)=P$, we obtain Eq. (\ref{eq40}). \qed
\vskip0.2in

\noindent {\bf 4. Coherence and decoherence in noisy Shor's algorithm}

In the real world, quantum systems cannot be completely isolated from their environment and are subject to various noise influences. Furthermore, quantum systems inevitably interact with their surrounding environments, leading to decoherence phenomena. In this section, we investigate coherence and decoherence in noisy Shor's algorithm.

Now we consider an important channel $\Phi$ acting on a
$d$-dimensional quantum system called the depolarizing
channel\cite{HXF,GYH,LLS19}
\begin{equation*}
\Phi(\rho) = \lambda V\rho V^{\dagger} + (1-\lambda)\frac{\mathbf{I}_{AB}}{d}, \quad \frac{-1}{d^2-1} \leq \lambda \leq 1, \quad
\end{equation*}
where $V$ represents an arbitrary unitary operator. The parameter
$\lambda$ is chosen to ensure that the channel $\Phi$ maintains
complete positivity, which is essential for physical realizability.
Since all depolarizing channels satisfy the unitality
($\Phi(\mathbf{I})=\mathbf{I}$), the set of depolarizing channels
constitutes a proper subset of the broader class of unital channels.
When $\lambda=0$, $\Phi$ reduces to the completely depolarizing
channel, which maps any input state to the maximally mixed state
$\frac{\mathbf{I}_{AB}}{d}$.

The depolarizing channel exhibits the fundamental
property of unitary invariance and constitutes a significant
subclass of commutativity-preserving channels. This channel
possesses the remarkable characteristic of never increasing
quantumness between any two quantum states. When $\lambda \neq 0$,
the depolarizing channel stands as one of the few channel types
incapable of creating quantum discord in zero quantum discord
states. Moreover, for quantum systems with dimension $n \geq 3$, it
occupies a privileged position due to its unique bidirectional
property of preserving quantum correlations, distinguishing it from
other quantum channels in the context of quantum information
theory.


Here, we discuss the case $V=W_A\otimes \mathbf{I}_B$ where $W_A=\sum_{x}\mathrm{e}^{\mathrm{i}\theta_{x}}|x\rangle\langle x|$ is an arbitrary diagonal unitary matrix. This specific scenario allows us to examine the properties and implications of diagonal unitarity within quantum computational frameworks. Let $\Phi$ be the noise channel. Then the noisy Shor's algorithm
can be described as
\begin{equation}\label{eq41}
\mathcal{E}=(\mathcal{F}^{\dagger}\otimes \mathbb{I})\circ\Phi\circ \mathcal{U}\circ\Phi,
\end{equation}
where $\Phi(\rho) = \lambda (W_A\otimes \mathbf{I}_B)\rho (W_A\otimes \mathbf{I}_B) + (1-\lambda)\frac{\mathbf{I}_{AB}}{d}$, $(\mathcal{F}^{\dagger}\otimes \mathbb{I})(\rho)=(F^{\dagger}_A\otimes \mathbf{I}_B)\rho (F_A\otimes \mathbf{I}_B)$ and $\mathcal{U}=U\rho U^{\dagger}$.
By applying the Hadamard gate, we obtain the initial state $|\psi_{1}\rangle=\frac{1}{\sqrt{Q}}\sum_{j=0}^{Q-1}|j\rangle_A|1\rangle_B$. Subsequently, applying the operator $W_A\otimes \mathbf{I}_B$ to this state yields
\begin{equation*}
(W_A\otimes \mathbf{I}_B)|\psi_1\rangle = \frac{1}{\sqrt{Q}}\sum_{j}\mathrm{e}^{\mathrm{i}\theta_{j}}|j\rangle_A |1\rangle_B.
\end{equation*}
Next, we consider the unitary transformation $U$ acting on the
previously obtained state, and obtain
\begin{equation*}
|\psi'_2\rangle=U(W_A\otimes \mathbf{I}_B)|\psi_1\rangle = \frac{1}{\sqrt{Q}}\sum_{j}\mathrm{e}^{\mathrm{i}\theta_{j}}|j\rangle_A |x^j \bmod N\rangle_B.
\end{equation*}
By imposing the operator $W_A\otimes \mathbf{I}_B$ on the state
$|\psi'_2\rangle$, we obtain
\begin{equation*}
(W_A\otimes \mathbf{I}_B)U(W_A\otimes \mathbf{I}_B)|\psi_1\rangle= \frac{1}{\sqrt{Q}}\sum_{j}\mathrm{e}^{2\mathrm{i}\theta_{j}}|j\rangle_A |x^j \bmod N\rangle_B.
\end{equation*}
Finally, we apply the inverse Fourier transform $F^\dagger_A\otimes
\mathbf{I}_B$, and get
\begin{equation*}
|\psi'_3\rangle=(F^\dagger_A\otimes \mathbf{I}_B)(W_A\otimes \mathbf{I}_B)U(W_A\otimes \mathbf{I}_B)|\psi_1\rangle= \frac{1}{Q}\sum_{j,k}\mathrm{e}^{2\mathrm{i}\theta_{j}}\mathrm{e}^{\frac{-2\pi\mathrm{i}jk}{Q}}
|k\rangle_A |x^j \bmod N\rangle_B.
\end{equation*}
Then the state after noisy Shor's algorithm can be described as
\begin{align*}
\mathcal{E}(|\psi_1\rangle\langle\psi_1|)
&= (\mathcal{F}^{\dagger}\otimes \mathbb{I})\circ\Phi\circ\mathcal{U}\left(\lambda (W_A\otimes \mathbf{I}_B)|\psi_1\rangle\langle\psi_1|(W^{\dagger}_A\otimes \mathbf{I}_B) + (1-\lambda)\frac{\mathbf{I}_{AB}}{d}\right)\notag\\
&= \lambda^2|\psi'_3\rangle\langle\psi'_3| + (1-\lambda^{2})\frac{\mathbf{I}_{AB}}{d}.
\end{align*}
{\bf Theorem 6} For the noisy Shor's algorithm caused by the
depolarizing channel where $W_A=\sum_{x}\mathrm{e}^{\frac{\mathrm{i}\pi
x}{Q}}|x\rangle\langle x|$, the coherence of state $\sigma_{3}$ with
respect to the von Neumann measurement $\Pi$ is quantified by
\begin{equation}\label{eq42}
C_{f}(\sigma_{3}, \Pi)= \frac{f(0)\lambda^4}{2}\left(1-\frac{1}{r^{2}}\right) \left[ \frac{d}{(1-\lambda^2) \cdot f\left(\frac{\lambda^2(d-1)+1}{1-\lambda^2}\right)} + \frac{d}{(\lambda^2(d-1)+1) \cdot f\left(\frac{1-\lambda^2}{\lambda^2(d-1)+1}\right)} \right],
\end{equation}
where $\frac{-1}{d^2-1} \leq \lambda< 1$ and $d=2^{n}$ is the
dimension of the composite system $AB$, and the decoherence of the
initial state $\rho_{1}$ induced by the quantum channel
$\mathcal{E}$ during noisy Shor's algorithm is characterized by
\begin{equation}\label{eq43}
D(\rho_{1},\mathcal{E})
=\lambda^{2}\left(\frac{1}{d}-\frac{1}{Q}\right)+\frac{d-1}{d}.
\end{equation}
$\it{Proof}$. Let
$\sigma_{3}=\mathcal{E}(|\psi_1\rangle\langle\psi_1|)$. The spectral
decomposition of the state $\sigma_{3}$ takes the form
\begin{equation*}
\sigma_{3} = \lambda_1|\psi'_3\rangle\langle\psi'_3| + \sum_{j=2}^d \lambda_j|\varphi''_j\rangle\langle\varphi''_j|,
\end{equation*}
where $\lambda_1= 1 - \frac{(d-1)(1-\lambda^{2})}{d}$ and $\lambda_j
= \frac{1-\lambda^{2}}{d}$ $(j\in \{2, 3, \ldots, d\})$. Here,
$\{|\varphi''_j\rangle\}_{j=1}^d$ represents the orthonormal basis,
where $|\varphi''_1\rangle=|\psi'_3\rangle$ and
$\langle\varphi''_j|\psi'_3\rangle = 0$ for $j \geq 2$. According to
Eqs. (\ref{eq8}) and (\ref{eq11}), the coherence of state
$\sigma_{3}$ with respect to the von Neumann measurement $\Pi$ is
\begin{equation*}
C_{f}(\sigma_{3}, \Pi)= \sum_i F_f(\sigma_{3}, K_i) = \frac{f(0)}{2} \sum_{jk} \frac{(\lambda_j - \lambda_k)^2}{\lambda_k f(\lambda_j/\lambda_k)} \langle\varphi''_j|\Pi(|\varphi''_k\rangle\langle\varphi''_k|)|\varphi''_j\rangle.
\end{equation*}
Utilizing the relation $\sum_{j=2}^d
|\varphi''_j\rangle\langle\varphi''_j| = \mathbf{I}_{AB} -
|\psi'_3\rangle\langle\psi'_3|$, after algebraic manipulation, the
value of $C_{f}(\sigma_{3}, \Pi)$ is
\begin{equation*}
\frac{f(0)\lambda^4}{2} \left[ \frac{d}{(1-\lambda^2) \cdot f\left(\frac{\lambda^2(d-1)+1}{1-\lambda^2}\right)} + \frac{d}{(\lambda^2(d-1)+1) \cdot f\left(\frac{1-\lambda^2}{\lambda^2(d-1)+1}\right)} \right]\left( 1 - \sum_{i=0}^{d-1} |\langle i|\psi'_3\rangle|^4 \right),
\end{equation*}
where $\lambda\neq1$ and $\langle i|$ denotes $\langle i|_{AB}$ for
the sake of simplicity. To simplify the analysis, we set $Q=rm$
where $m$ is a positive integer, which implies that
$\left\lfloor\frac{Q-1}{r}\right\rfloor = m-1$. Decomposing the
index $j$ as $j = a + br$, where $a$ is the remainder and $b$ is the
quotient, we can rewrite state $|\psi'_3\rangle$ as
\begin{equation*}
|\psi'_3\rangle=\frac{1}{Q}\sum_{k=0}^{Q-1}\sum_{a=0}^{r-1}\sum_{b=0}^{m-1}\mathrm{e}^{2\mathrm{i}\theta_{a+br}}
\mathrm{e}^{\frac{-2\pi\mathrm{i}(a+br)k}{Q}}|k\rangle_A |x^a \bmod N\rangle_B.
\end{equation*}
The coherence of $|\psi'_3\rangle$ with respect to the von Neumann
measurement $\Pi$ is $1 - \sum_{i=0}^{d-1} |\langle
i|\psi'_3\rangle|^4$. When $W_A=\sum_{x}\mathrm{e}^{\frac{\mathrm{i}\pi
x}{Q}}|x\rangle\langle x|$, we obtain
\begin{align*}
1 - \sum_{i=0}^{d-1} |\langle i|\psi'_3\rangle|^4
&=1-\frac{1}{Q^{4}}\sum_{k=0}^{Q-1}\sum_{a=0}^{r-1}\left|\sum_{b=0}^{m-1}
\mathrm{e}^{2\mathrm{i}\theta_{a+br}}
\mathrm{e}^{\frac{-2\pi\mathrm{i}kbr}{Q}}\right|^{4}\notag\\
&=1-\frac{1}{Q^{4}}\sum_{k=0}^{Q-1}\sum_{a=0}^{r-1}\left|\sum_{b=0}^{m-1}
\mathrm{e}^{\frac{-2\pi\mathrm{i}b(k-1)}{m}}\right|^{4}=1-\frac{1}{r^{2}}.
\end{align*}
Thus we can derive Eq. (\ref{eq42}). According to Eqs. (\ref{eq10})
and (\ref{eq14}), we have
\begin{equation*}
D(\rho_{1},\mathcal{E})
=\frac{1 + \text{tr}(\rho_{1} \mathcal{E}(\mathbf{I}_{AB}))}{2} -  \text{tr}(\rho_{1}\mathcal{E}(\rho))
=\frac{d-1+\lambda^{2}}{d}-\lambda^{2}\left|\langle\psi_1|\psi'_3\rangle\right|^2.
\end{equation*}
Direct calculations show that
\begin{equation*}
\left|\langle\psi_1|\psi'_3\rangle\right|^2 =\frac{1}{Q^{3}}\left|\sum_{k=0}^{Q-1}\sum_{b=0}^{m-1}\mathrm{e}^{2\mathrm{i}\theta_{br}}
\mathrm{e}^{\frac{-2\pi\mathrm{i}brk}{Q}}\right|^2
=\frac{1}{Q^{3}}\left|\mathrm{e}^{2\mathrm{i}\theta_{0}}Q\right|^2=\frac{1}{Q}.
\end{equation*}
Therefore,  Eq. (\ref{eq43}) holds. \qed

Taking $f(x) = \left(\frac{1 + \sqrt{x}}{2}\right)^2$, we immediately obtain the following result.\\
{\bf Corollary 3} For the noisy Shor's algorithm caused by the
depolarizing channel where $W_A=\sum_{x}\mathrm{e}^{\frac{\mathrm{i}\pi
x}{Q}}|x\rangle\langle x|$, the coherence of state $\sigma_{3}$ with
respect to the von Neumann measurement $\Pi$ is given by
\begin{equation}\label{eq44}
C(\sigma_{3}, \Pi)= \left(\sqrt{1 - \frac{(d-1)(1-\lambda^{2})}{d}} - \sqrt{\frac{1-\lambda^{2}}{d}}\right)^2\times\left(1-\frac{1}{r^{2}}\right).
\end{equation}

Since $\frac{-1}{d^2-1} \leq \lambda  \leq 1$, we can obtain $1 - \frac{1}{Q} \leq D(\rho_{1},\mathcal{E}) \leq 1-\frac{1}{d}$ and $0 \leq C(\sigma_{3}, \Pi) \leq 1-\frac{1}{r^{2}}$.
When $\lambda=0$, $\Phi$ reduces to the completely depolarizing channel, which maps any input state to the maximally mixed state $\frac{\mathbf{I}_{AB}}{d}$, resulting in the complete loss of information from the original state. In this scenario, the decoherence induced on the state $|\psi_1\rangle$ by the channel $\mathcal{E}$ reaches its maximum value, quantified as $D(\rho_1,\mathcal{E}) = 1 - \frac{1}{d}$. However, the coherence of state $\sigma_{3}$ with respect to $\Pi$ reaches its minimum value of 0 when $W_A=\sum_{x}\mathrm{e}^{\frac{\mathrm{i}\pi x}{Q}}|x\rangle\langle x|$.

When $\lambda=1$, $\Phi$ reduces to a unitary
channel, which merely performs a unitary transformation while
perfectly preserving the quantum state information. Under these
conditions, the decoherence induced on the state $|\psi_1\rangle$ by
the channel $\mathcal{E}$ attains its minimum value, represented as
$D(\rho_1,\mathcal{E}) = 1 - \frac{1}{Q}$. When
$W_A=\sum_{x}\mathrm{e}^{\frac{\mathrm{i}\pi x}{Q}}|x\rangle\langle
x|$, the coherence of state $\sigma_{3}$ with respect to $\Pi$
reaches its maximum value of $1-\frac{1}{r^{2}}$. Furthermore,
$\lambda = -\frac{1}{d^2 - 1}$ represents the critical parameter
value for maintaining complete positivity of the channel. Any value
of $\lambda$ smaller than this threshold would cause the channel to
lose its complete positivity property. At this critical point, the
decoherence and coherence are given by $D(\rho_1,\mathcal{E}) =
\frac{1}{(d^2 - 1)^2} \times (\frac{1}{d} - \frac{1}{Q}) +
\frac{d-1}{d}$ and $C(\sigma_{3}, \Pi) = \left(\sqrt{1 -
\frac{d(d-1)(d^2-2)}{(d^2-1)^2}} -
\sqrt{\frac{d(d^2-2)}{(d^2-1)^2}}\right)^2 \times
\left(1-\frac{1}{r^{2}}\right)$.

Since the eigenvectors of $U_{B}$ are $|u_{s}\rangle$, $|\psi'_2\rangle$ can be rewritten as
\begin{align}\label{eq45}
U\frac{1}{\sqrt{Q}}\sum_{j=0}^{2^{t}-1}\mathrm{e}^{\mathrm{i}\theta_{j}}|j\rangle_A |1\rangle_B
&= \sum_{j=0}^{2^{t}-1}\frac{1}{\sqrt{Q}}\mathrm{e}^{\mathrm{i}\theta_{j}} |j\rangle_A U_{B}^{j}\left(\frac{1}{\sqrt{r}} \sum_{s=0}^{r-1} |u_s\rangle_B\right)\notag\\
&= \frac{1}{\sqrt{Qr}} \sum_{j=0}^{2^{t}-1}\sum_{s=0}^{r-1} \mathrm{e}^{\mathrm{i}\theta_{j}}\mathrm{e}^{2\pi \mathrm{i}j \frac{s}{r}} |j\rangle_A |u_s\rangle_B.
\end{align}
We then apply the operator $W_A\otimes \mathbf{I}_{B}$ and the inverse Fourier transform $F^\dagger$, $|\psi'_3\rangle$ can be rewritten as
\begin{align}\label{eq46}
&(F^\dagger_A \otimes \mathbf{I}_B)\left(\frac{1}{\sqrt{Qr}} \sum_{j=0}^{2^{t}-1}\sum_{s=0}^{r-1} \mathrm{e}^{2\mathrm{i}\theta_{j}}\mathrm{e}^{2\pi \mathrm{i}j \frac{s}{r}} |j\rangle_A |u_s\rangle_B\right)\notag\\
&= \frac{1}{Q\sqrt{r}} \sum_{j,k=0}^{2^{t}-1}\sum_{s=0}^{r-1} \mathrm{e}^{\mathrm{i}\theta_{j}}\mathrm{e}^{2\pi \mathrm{i}j (\frac{s}{r}-\frac{k}{Q})}|k\rangle_A|u_s\rangle_B.
\end{align}
When measuring register $A$, the probability of obtaining outcome $k$ is
\begin{equation}\label{eq47}
P_{\lambda}(k)=\sum_{j=0}^{2^{n-t}-1}\left\langle k,j\left|\left[\lambda^2|\psi'_3\rangle\langle\psi'_3| + (1-\lambda^{2})\frac{\mathbf{I}_{AB}}{d}\right]\right|k,j\right\rangle.
\end{equation}
\noindent{\bf Theorem 7} When $W_A=\sum_{x}\mathrm{e}^{\frac{\mathrm{i}\pi x}{6Q}}|x\rangle\langle x|$, the probability that we can calculate $r$ in noisy Shor's algorithm is
\begin{equation}\label{eq48}
P_{\lambda}> \frac{27\lambda^{2}\phi(r)}{16r\pi^2}+\frac{(1-\lambda^{2})\phi(r)}{Q}.
\end{equation}
$\it{Proof}$. Let
$P_{|\psi'_3\rangle}(k)=\sum_{j=0}^{2^{n-t}-1}\langle
k,j|\psi'_3\rangle\langle\psi'_3|k,j\rangle$. Through rigorous
calculations, we have
\begin{equation*}
P_{|\psi'_3\rangle}(k) =\frac{1}{Q^{2}r}\sum_{s=0}^{r-1} \left|\sum_{j=0}^{2^{t}-1}\mathrm{e}^{2\mathrm{i}\theta_{j}} \mathrm{e}^{2\pi \mathrm{i}j(\frac{s}{r}-\frac{k}{Q})}\right|^2.
\end{equation*}
When $W_A=\sum_{x}\mathrm{e}^{\frac{\mathrm{i}\pi x}{6Q}}|x\rangle\langle x|$, we have
$
P_{|\psi'_3\rangle}(k) =\frac{1}{Q^{2}r}\sum_{s=0}^{r-1} \left|\sum_{j=0}^{2^{t}-1} \mathrm{e}^{2\pi \mathrm{i}j(\frac{s}{r}-\frac{k}{Q}+\frac{1}{6Q})}\right|^2.
$
Denote $\sum_{j=0}^{2^{t}-1} \mathrm{e}^{2\pi \mathrm{i}j(\frac{s}{r}-\frac{k}{Q}+\frac{1}{6Q})}$ by $R$. Specifically, we have
\begin{equation*}
|R|^2 > \frac{27Q^2}{16\pi^2} \quad \text{for} \quad \left|\frac{s}{r}-\frac{k}{Q}\right| < \frac{1}{2Q}.
\end{equation*}
According to Eq. (\ref{eq47}), we obtain that the probability of obtaining outcome $k$ is
\begin{equation*}
P_{\lambda}(k) = \lambda^2P_{|\psi'_3\rangle}(k)+\frac{1-\lambda^{2}}{Q}.
\end{equation*}
For any given ratio $\frac{k}{Q}$, there exists a unique fraction $\frac{s}{r}$ with $r < N$ satisfying this condition. Consequently, the probability that $\frac{k}{Q}$ serves as the optimal approximation to a fraction with denominator $r$ satisfies
\begin{equation*}
P'_{\lambda}(k)>\frac{27\lambda^{2}}{16r\pi^2}+\frac{1-\lambda^{2}}{Q}.
\end{equation*}
The algorithm requires that the numerator $s$ be coprime to $r$, preventing cancellation of common factors that would otherwise reduce the fraction $s/r$. Then the probability of successfully calculating $r$ becomes
$
P_{\lambda}> \frac{27\lambda^{2}\phi(r)}{16r\pi^2}+\frac{(1-\lambda^{2})\phi(r)}{Q}.
$
\qed\\
{\bf Remark 3} When $W_A=\sum_{x}\mathrm{e}^{\frac{\mathrm{i}\pi
x}{Q}}|x\rangle\langle x|$, the probability of obtaining outcome $k$
with decoherence of the initial state $\rho_{1}$ satisfies the
following relation
\begin{equation}\label{eq49}
D(\rho_{1},\mathcal{E})-P_{\lambda}(k)
=\gamma(\lambda),
\end{equation}
where
$\gamma(\lambda)=\lambda^{2}(\frac{1}{d}-\frac{1}{r})+1-\frac{1}{d}-\frac{1}{Q}$.
Since $\frac{-1}{d^2-1} \leq \lambda  \leq 1$, we obtain $1 -
\frac{1}{r} - \frac{1}{Q} \leq
D(\rho_{1},\mathcal{E})-P_{\lambda}(k) \leq 1 - \frac{1}{d} -
\frac{1}{Q}$.

\vskip0.1in

\noindent {\bf 5 Conclusions and discussions}\\\hspace*{\fill}\\
In this study, we have systematically analyzed the coherence and decoherence dynamics in Shor's algorithm under both noiseless and noisy conditions, and have established lower and upper bounds for the probability of successfully determining the order $r$.
For the generalized Shor's algorithm, we have shown the upper bound of the probability that we can calculate $r$ exhibits a negative correlation with coherence of state $\rho'_{3}$ relative to the von Neumann measurement. The lower bound of the probability that we can calculate $r$ increases with increasing $Q$.
When the register is initialized in a pseudo-pure state, we have investigated the coherence and decoherence dynamics in Shor's algorithm. For large values of $Q$, the probability that we can calculate $r$ when the register is initialized in a pseudo-pure state is approximately proportional to the probability that we can calculate $r$ when the register is initialized in an arbitrary pure state.

In addition,  we have investigated coherence and
decoherence dynamics in noisy Shor's algorithm. It has been revealed
that while the coherence of state $\sigma_3$ relative to the von
Neumann measurement depends on the order $r$, the decoherence
induced on state $|\psi_1\rangle$ by the quantum channel
$\mathcal{E}$ remains independent of $r$. For the specific case
where $W_A=\sum_x \mathrm{e}^{\frac{i\pi x}{6Q}}|x\rangle\langle
x|$, we have rigorously proven that the lower bound of the success
probability is determined by the parameters $Q$, $r$, and $\phi(r)$,
where $\phi(r)$ denotes Euler's totient function. Our results may
offer new insights into the study of resources in quantum
algorithms.

In \cite{SY}, the authors investigated the Grover
search algorithm with an arbitrary pure initial state or in a
pseudo-pure state, and showed that the coherence, quantified by the
Wigner-Yanase skew information, satisfies a complementary relation
or an approximate one with the success probability. Moreover, the
relation between decoherence and success probability under
single-qubit noise was analyzed. In this work, we shift our focus on
Shor's algorithm in both pure state and pseudo-pure state case.
Instead of deriving complementary relations, we have deduced both
upper and lower bounds on the success probability by employing
metric-adjusted skew information of coherence, a class of more
general coherence quantifiers, which can also be interpreted as a
measure of quantum uncertainty or asymmetry in the state-channel
interplay\cite{LuoS2018}. This highlights the generality and
applicability of our approach. In addition, we obtain a lower bound
on the success probability under depolarizing channels in $d$
dimensional systems.

The Grover's algorithm was also considered in the
arbitrary pure initial state scenario, and the average success
probability is shown to exhibit a strict linear dependence on the
coherence fraction of the initial state, while the maximum coherence
fraction corresponds to the highest success probability\cite{ZSQ}.
We found that the probability of obtaining the outcome $k$ in
generalized Shor's algorithm reaches the maximum when the algorithm
reduces to the original one, where similar arguments has been
elucidated for Gover's algorithm in \cite{ZSQ}. 

We restrict attention to pseudo-pure states and our
main conclusions rely on the form of them. For general mixed input
states, quantifying coherence in Shor's algorithm becomes
computationally intractable and analytically challenging due to the
algorithm's intrinsic complexity, which makes a systematic
characterization of performance difficult. It is revealed that under
a fixed admixture, purer inputs tend to yield higher success
probabilities. The structured mixed inputs have been considered and
the robustness of coherence has been related to the performance of
the probabilistic BV algorithm\cite{NMTV}. However, most quantum
algorithms are designed to operate on pure input states.

\vskip0.1in

\noindent

\subsubsection*{Acknowledgements}
\small {This work was supported by National Natural Science
Foundation of China (Grant Nos. 12561084, 12161056); Natural Science Foundation of Jiangxi Province (Grant No. 20232ACB211003).}


\end{document}